# Jets and Accretion Disks in Astrophysics - A brief review

Linda A. Morabito[1] and David Meyer[1]

[1]Department of Astronomy, Victor Valley College, Victorville, CA 92395

linda.morabito@vvc.edu

Abstract

The significance of jets and accretion disks in Astrophysics may be growing far beyond any single example of recent finds in the scientific journals. This brief review will summarize recent, significant manifestations of accretion disk powered jets in the universe. We then introduce supplemental contemporary finds in physics and astrophysics which might bear tangential or direct implications for astrophysics toward rethinking the universe with a major role of relativistic jets powered by accretion disks. We conclude with the direction our research will take in order to establish a new perspective on the universe.

1. Introduction

The existence of accretion disks around many objects found in the universe, with charged particles accelerated to nearly the speed of light along twisted magnetic field lines or relativistic jets are ubiquitous in connection with the formation and death of stars. New evidence points to a similar role of supermassive black holes in regards to star formation from the perspective of a galaxy's life cycle. This mechanism of accretion disk powered relativistic jets may have even greater meaning when viewed in the context about their role in future research about the greater universe.

2. Jets and Accretion Disks in Brief Review

Recently it has been proposed that even Sun-like stars exhibit a stage as preplanetary nebulae when in binary systems, which produces jets along the star's poles in order to expel material (Kwok 2000). Stars produce jets during formation as they form from infalling material from an accretion disk (Reipurth 2001); neutron stars in binary systems become particular types of pulsars via the concentrated jets of material (Kaspi 2010); stars which died as black holes feed from accretion disks when in binary systems (Shakura 1973).

Common to jets and other stellar outflows are destructive factors in destruction of the molecular clouds from which the star itself was formed, thereby preventing the formation of more stars (Arce 2003). They also distribute material necessary for the production of future generations of stars, in that material with metals is necessary for planetary formation and life (Heger 2003). Jets from stars can send shock waves which compress regions of gas leading to the formation of more stars (Lee 2005). Jets can distribute material within their own system during a star's and surrounding planetary system's formation, while accomplishing the aforementioned (Ciesla 2009).

It has recently been suggested that on a larger scale, supermassive black holes in the centers of galaxies may form directly from cloud collapse (Begelman 2006). They may also remove material from a galaxy when they are active, heating gas and preventing its return to the galaxy to be used for further star formation, for long periods of time (Hota 2012; Page 2012; Tombesi 2012). Active jets are constituted when a supermassive black hole is actively feeding from an abundance of material in an accretion disk around it (McKinney 2012). Additionally, jets from supermassive black holes can promote star formation in intergalactic clouds with which they interact, thereby forming stars (van Breugal 2004; Croft 2006), promote star formation within their own galaxies (Crockett 2012), may be responsible for the formation of stars in early galaxies (Klamer 2004; Eisenhardt 2012), as well as distribute metals to other galaxies (Kirkpatric 2011).

Relativistic jets powered by the conversion of gravitational potential energy into kinetic and then thermal energy (powered by the conversion of energy within accretion disks), overall, can be summarized to yield the following results from interactions with the environment around them: 1) they prevent star formation; 2) they aid in star formation: and 3) therefore because of the previous two characteristics regulate star formation. A summary of tangential or direct related research results is also presented here:

## 3. Relevant Research in Physics and Astrophysics

Recent finds in physics and astrophysics may either tangentially or directly arise in the consideration of the significance of the role the formation of accretion disks and jets may play in astrophysics. By summarizing them here it is possible that further and insightful directions for research will result.

As per section 2 above, the mechanism of an accretion disk powered relativistic jet comes into existence also coupled with the formation of stars, and at the end of the lifetimes of stars in some cases, and in the case of supermassive black holes not only the beginning of their existence possibly within super-luminous galaxies from the early universe (Eisenhardt 2012), but also at the end of the life cycle of galaxies in regards to star formation (Kassin 2012; Page 2012). Additionally, Kassin (2012) establishes that spiral galaxies emerged from a settling or hydrodynamic process which defines galaxy life cycles. If accretion disks and relativistic jets can be viewed as ubiquitous mechanisms for change which may in fact be correlated to the beginning and ending of the life cycles of stars and galaxies, in a hierarchical view of the universe, previously discarded cosmological views based on energetic events of this nature at the beginning and ending of a hypothetical universe's "life cycle" may be re-examined in light of this observational evidence.

First, the nature of black holes has been under investigation in astrophysics for decades. Stephen Hawking concedes that information is not lost within a black hole and from his and the standpoints of many derived models of the universe, black holes do not collapse into a singularity when viewed for quantum mechanical reasons in imaginary time (Hartle 1983); or that there are quanta of the smallest particle of spacetime which will not permit black holes to collapse to a singularity (Ashtekar 2009); or that there are reasons based on a repulsion of antimatter to matter which prevent the collapse in a black hole to reach a singularity (Hajdukovic 2011a).

The ubiquitous nature of black holes in the universe when viewed in light of the aforementioned mechanism of change in the universe might be suggestive of a fresh view of black holes which do not in a practical sense lead anywhere beneath their event horizons (Hawking 2005), but rather are viewed as a repository for matter capable of yielding under a wide spectrum of circumstances these formations to influence their environment. When as mechanisms in the universe, which are powered by gravitational potential energy converted to heat in an accretion disk which surrounds them producing jets they accomplish change in the universe as previously described. It has been hypothesized that these mechanisms do not last forever, and that when not in use they do ultimately deconstruct themselves through Hawking radiation (Hawking 1974) or in some cases more rapidly through other means (Hajdukovic 2011b).

As mentioned it is possible that a hierarchical view of these observations could lead to a second look at cosmologies that predict a collapsing universe. As an example, Hajdukovic (2012a) derives a cosmological scale factor as a linear function of time in the late-time universe (Hajdukovic 2012a), which would permit a collapsing universe. Recent discoveries have researchers exploring dark energy as potentially a form of quiescence or zero point energy (Copeland 2006). Interestingly Hajdukovic (2012a) has explored dark energy in the context of a fluid of gravitational dipoles in the quantum vacuum as described by Hajdukovic (2012a). Additionally, the quest for reasons why the graviton exchange is a factor of approximately $10^{-40}$ the strength of the photon, gluon, or weak bosons exchange is a quest which could be answered by the superposition of a stronger manifestation of gravity caused by the alignment of gravitational dipoles, virtual particles from the quantum vacuum in the presence of matter (Hajdukovic 2011c), and specifically not a particle. Hajdukovic's (2011a, 2011c, 2012a) is a particularly poignant example of new research in physics because it has recently been predicted that antigravity from the quantum vacuum in the presence of baryonic matter could induce precession of the perihelion of Dysnomia's orbit around dwarf planet Eris (Hajdukovic 2012b).

Additionally, although (Hajdukovic 2011a) predicts a cyclic universe resulting from a Big Bang from the conversion of antimatter to matter (or matter to anti-matter) initiated by a Schwinger mechanism inside the black hole of a collapsed universe (the loss of mass of the black hole is equal to the amount of matter or antimatter taken from the quantum vacuum which is ejected from the black hole due to antigravity), a hierarchical extrapolation of the recent observational evidence in regards to accretion disks has intriguing possibilities such as collision produced pair production from jets (Stern 2006) or jet delivered (Williams 2004) matter antimatter pairs. Is it possible that it is time to reconstruct the physics of a collapsing universe in light of the possibility of the formation of an accretion disk before all the mass of a contracting universe is beneath the event horizon? The power of these jets as the mass of the universe tries to enter such a hypothetical black hole, including the possible feedback limitations on the size of a contracted universe's black hole, might produce profound considerations that may have been previously overlooked.

Finally, if the summary of accretion disks and jets which began this brief review correlates highly to regulation of star formation or creating conditions which spawn star formation or restraining those conditions, in essence one can interpret their effects as manifesting the rate of supernova explosions throughout the universe. In a recent paper Svensmark (2012) draws a strong correlation to the rate of nearby supernova explosions to the biodiversity rate on Earth. In essence a certain range of rate of supernova explosions is required to ensure the flourishing of life as we know it, with none too close, and not exceeding that range. Is it possible that the arising of high-mass settled disk spiral galaxies in the universe, a process which began after the rate of galaxy collisions and star formation began to drop off in the universe at $z \sim 1$ (Kassin 2012), produced conditions for the emergence of civilizations like ours? There have apparently been perhaps as many as 4.6 billion years of conditions where the supernova rate near Earth was conducive to our development, and perhaps the remaining lifetime of our Sun or our galaxy's collision with another massive spiral galaxy before that stability would end.

4. Toward a New View of the Universe

Recently it has been suggested that the outflow power of black hole winds and jets can be described as a function of the Eddington fraction directly associated with black hole accretion (King 2012). This is an extremely important result because it has been derived across the mass scale from black hole binaries (BHB) to supermassive black holes (SMBH) of $10^8$ solar masses. The results of this study find that the mechanism that regulates winds is a simple relation over a large mass scale, and the study data may be consistent with winds and jets being regulated in a common fashion (King 2012). It is too early to say but these results may be consistent with a hierarchical perspective of the universe, as described in section 3 above, which ascribes common processes to varying scales of stellar mass black holes in binaries and supermassive black holes in the centers of galaxies without dependence on black hole size.

The goal of our research will be to determine if it is possible to describe and constrain a viable progenitor universe to evolve, as do smaller structures within the hierarchy. If so, the description of this progenitor universe would serve to narrow the range of cosmological models that support its existence. A viable cosmological model would not only permit our universe to be consistent within its hierarchy with the origin and fate of the smaller scale structures in our universe, but also with the arising of an accretion disk and bipolar jets in conjunction with its formation and then final stage. Such a cosmology would also allow our universe to be consistent with a generational perspective now ascribed to stars and to galaxies. This perspective would not support continued accelerating expansion of the universe, and is possibly inconsistent with a cyclic universe, which arises in a sense as a direct replacement of its predecessor.

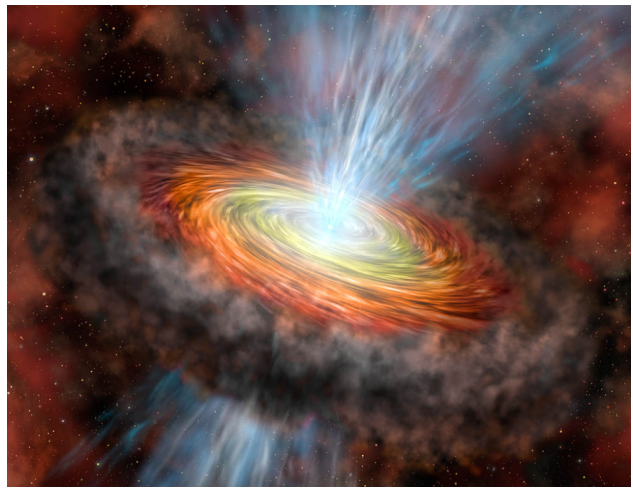

Might the birth process of a star give clues as to the birth of our universe?
"Gemini Observatory/Association of Universities for Research in Astronomy"

There are two further implications of the analysis by King (2012) we believe may have bearing on our own direction. More massive black holes may actually be more efficient in expelling material than less massive ones, and the postulated congruence of the manner of the regulation of winds and jets to one another. Magneto-hydrodynamics therefore might be the process for the black hole winds, since they are likely needed to account for the enormous energy of jets composed of particles accelerated to nearly the speed of light.

Very recent breakthroughs in magneto-hydrodynamic modeling of jets suggest that for a magnetically-arrested disk (MAD) outflow efficiency is solely dependent upon spin rate of the black hole and the angular density thickness of the accretion flow h/r (Tchekhovskoy 2012). According to the study, the efficiency of some jets and winds, given by the ratio of their combined power to the amount of rest mass energy supplied by the accretion can yield efficiencies even several times greater than 100% for $\eta = (P_{jet} + P_{wind})/\langle Mc^2 \rangle \times 100\%$. Thus efficiency is maximized for any specific amount of spin and disk thickness. For varying amounts of black hole spin, we will consider the type of angular density thickness we would expect to find in a collapsing progenitor universe before accretion is shut down in the progenitor universe's black hole (via accumulation of magnetic flux (Tchekhovskoy 2012)).

There is further possibly of uncovering an entirely new paradigm that has never before been considered in the context of cosmology or black hole formation, which we will explore because it may yield other universes. We would want to derive the nature of the environment surrounding a progenitor universe as it collapses, once we have established the power of the jets that have emerged from the progenitor's black hole. This previously unmodeled environment of debris above the event horizon of the magnetically-arrested progenitor's black hole might be triggered into the formation of other universes; perhaps one or two universes arising sequentially from the respective shock waves of the bipolar jets of the progenitor universe. The nature of this environment could open up new possibilities as to how our universe might have arisen as well as provide further insight into the processes we observe lower in the hierarchy. The problem could be constrained from either end; the determination of the mass outside an event horizon when the progenitor universe might prevent continued black hole growth through collapse, or the power needed to interject material and collapse into a previously unmodeled environment to yield another universe, which would give clues to the size of the black hole portion of the progenitor universe when feedback may have prevented its growth.

5. In Summary

We have shown interest in the work of Hajdukovic (2012a) in regards to cosmology on the basis that it predicts a collapsing universe built on a premise of gravitational repulsion of matter to antimatter in the quantum vacuum. In addition, the effects of dark matter arise in Hajdukovic's work from the polarization of the quantum vacuum in the presence of matter, rather than through a particle that has not yet been found. Although among the work's predictions is a cyclic universe, the mechanism of the rapid decrease in the mass of the black hole to spawn a new universe directly from the progenitor universe may be linked to a gravitational field which requires greater strength than is generated in the progenitor black hole. If so, this may not place the body of this work at odds to our hierarchical approach.

We will explore a limiting mass of a progenitor universe's central black hole in the framework of the results of recent magneto-hydrodynamic simulations that appear to match observations of active galactic nuclei. Since we are at the onset of this work, we have been greatly encouraged by the finding of regulation in a common fashion of the power of black hole jets and winds over the large range of black hole masses considered (King 2012), which might be consistent with our hierarchical approach.

Furthermore, we would also be greatly encouraged to find, enhance, or develop a cosmology consistent with our results that had testable predictions. It could be reasonably stated that our goal is to unify physics with astrophysics by leading toward cosmology which is founded on a physical property of the universe such as the nature of the quantum vacuum, rather than on mathematical derivation such as string theory. Simply put we seek a perspective of the universe that can be tested as Einstein's General Relativity was tested through observation of the advancement of the perihelion of Mercury. Should the gravitational nature of the quantum vacuum be indirectly supported by our work, observations of the retrograde perihelion precession of Dysnomia around dwarf planet Eris one day – a prediction of antimatter gravity in the quantum vacuum (Hajdukovic 2012b) – will likely yield no less excitement than Mercury in that day, and no less a breakthrough in our understanding of the universe.

Our work to derive a universe consistent with recent observations and breakthroughs in interpretation of the life cycles of stars and galaxies, as the highest-level hierarchy we encounter may give us the ability to appreciate the progenitor universe which helped our universe arise, and understand how these processes may have laid the foundation for stable environments conducive to our development.